# Improved full-waveform inversion for seismic data in the presence of noise based on the K-support norm

Jiahang Li[1], Hitoshi Mikada[1], and Junichi Takekawa[1]

[1]Department of Civil and Earth Resources Engineering, Kyoto University

**Running head:** FWI based on the K-support norm

## ABSTRACT

In the inversion part of the full-waveform inversion (FWI) that brings high resolution in finding a convergence point in the model space, a local numerical optimization algorithm minimizes the objective function based on the $\ell_2$ norm using the least-square form. Since the $\ell_2$ norm is sensitive to outliers and noise, the method may often lead to inaccurate imaging results. Thus, a new regulation form with a more practical relaxation form is proposed to solve the overfitting drawback caused by the use of the $\ell_2$ norm, namely the K-support norm, which has the form of more reasonable and tighter constraints. In contrast to the least-square method that minimizes the $\ell_2$ norm, $\ell_1$ norm can guarantee sparsity and robustness, as well as excellent denoising ability; thus, we consider a new regularization form, the K-support norm, which combines the $\ell_2$ and the $\ell_1$ norms in minimization. Then, a quadratic penalty method, which can speed up the convergence rate and avoid falling into local minima, is adopted to linearize the non-linear problem to lighten the computational load. This paper introduces the concept of the K-support norm and integrates this scheme with the quadratic penalty problem to improve the convergence and robustness against background noise. In the numerical example, three synthetic models are tested to clarify the effectiveness of the K-support norm by comparison to the Tikhonov regularization with a noisy data set. We considered two types of noise, i.e., random and coherent noise. Experimental results indicate that the modified FWI based on the new regularization form effectively improves inversion accuracy and stability, which significantly enhances the lateral resolution of depth inversion even with data with a low signal-to-noise ratio (SNR).

Full-waveform inversion (FWI) is a process of fitting seismic wavefields developed from seismic survey inversion techniques (Sirgue and Pratt, 2004). In the optimization process, the traditional FWI calculates the difference between the recorded and the simulated wavefields in the criterion of least-square methods (Lailly et al., 1983; Tarantola, 1984). For finding the convergent point in the model space, the $\ell_2$ norm using the difference between the recorded and the simulated wavefields in the objective function is minimized. The best fit of the model to the recorded seismic data (Virieux and Operto, 2009) is, thus, achieved at the convergent point that is a typical solution in the application of the least square method in geophysics (Choir and Alkhalifah, 2012).

However, the error contained in the observed data could be amplified when high non-Gaussian white noise or outliers are included (Brossier et al., 2007) due to the error being squared in the $\ell_2$ norm evaluation. We can understand this phenomenon as the weight decay feature of the $\ell_2$ norm, which amplifies the error, making the $\ell_2$ norm-based FWI more sensitive to models inferred from data with outliers (Zhong and Zhang, 2013). In summary, under high background noise conditions, the conventional $\ell_2$ norm may not sufficiently get rid of the negative impact of outliers and noise in the inverse procedure. Therefore, a new norm form with sparsity and a better and more appropriate constraint form should be applied to the modified FWI.

The sparse constraint form with good robustness is the $\ell_1$ norm (Schmidt, 2005), which is relatively insensitive to noise and outliers in seismic data processing (Kwak, 2008). The $\ell_1$ norm is often used to feature the selection of sparse observation data for the forward solution in FWI, resulting in a sparse solution (Loris et al., 2007). This sparse solution means that in the simulated velocity wavefield, as the sparsity increases, the coefficients of more variables become zero. Once the coefficient of a variable is zero, the corresponding variable is forced to zero and has no longer an impact to the model, i.e., part of the data in the simulated data is forced to zero. FWI adopts such an approach, which reduces the complexity of the model, but valuable information can be retained, which is why the $\ell_1$ regularization can also prevent the model from overfitting (Koh et al., 2007). In other words, the result of sparsity can reduce the impact of outliers on the original data by controlling model complexity (Donoho, 2006). For example, when we minimize an objective function, the $\ell_1$ norm will encourage the sparsity of its residuals. It mainly restricts the model parameters through the regular term, making it as small as possible or as close to zero as possible so as to achieve the effect of sparse model parameters, thereby reducing the complexity of the model. In addition, the $\ell_1$ norm can make some valuable components in the residual very large, which can avoid many small components (outliers and noise) interfering with the residual value. A more popular understanding is that the $\ell_1$ norm can retain useful velocity features and ignore those small anomalous disturbance components, and these small components are often the background noise we want to remove from the data (Boyd and Vandenberghe, 2004).

Based on the discussion above, we try to use a constraint form combining the $\ell_1$ and $\ell_2$ norms in FWI as a new objective function, namely K-support regularization. In mathematics, the K-support norm is first proposed; it has the characteristics of the tightest convex relaxation of sparsity and maintains the characteristics of the $\ell_2$ norm, which means this method maintains the sparsity and robustness of the $\ell_1$ norm while maintaining the weight decay properties of the $\ell_2$ norm (Bai and Liang, 2020). At the same time, the objective function still maintains convex, which ensures that the algorithm can find the globally optimal solution (McDonald et al., 2014).

By adjusting the control parameters, the K-support norm may be characterized as a particular norm intermediate between the $\ell_1$ and the $\ell_2$ norms (Haury, 2012). This approach can provide tighter constraints and relaxation of overfitting ability to help FWI based on the new regularization restrain the outliers and noise to a certain extent.

Since the wave equation itself is a non-linear partial differential equation, and the simulated data are implemented by solving the model parameters in the non-linear wave equation, the relationship between the observed and simulated data is also highly non-linear. This fact leads to the fact that the misfit function calculated in the least-squares form in FWI is also non-linear, and it becomes easy to fall into local minima. However, the quadratic penalty algorithm introduces a quadratic term in the loss function, which is equivalent to a restriction on the model parameter and reduces the nonlinearity in the inversion problem. Therefore, we propose to use the quadratic penalty term misfit function in FWI while combining the K-support norm to minimize the difference between the observed and simulated wavefields. In addition, contamination of large noise will lead to overfitting of the $\ell_2$ norm, which will reduce the inversion accuracy. In practical production, the information of noise is non-A Priori, so the modified algorithm has better flexibility and applicability than using only $\ell_1$ or $\ell_2$ norm. Thus, we no longer need to go through multiple tests before deciding which regularization to use. We only use the K-support norm, i.e., a more reasonable regularization with unknown noise information.

In this paper, based on the quadratic penalty problem scheme, we innovatively introduce the K-support norm algorithm to improve the resolution of FWI for data sets with high background noise. The acceleration in computation is realized by adopting the alternating direction method of multipliers (ADMM) algorithm. Three numerical models (Marmousi II, 2004 BP and SEG/EAGE Overthrust) are tested by comparing the Tikhonov regularization FWI with the K-support norm FWI to verify the superiority of our method.

van Leeuwen and Herrmann (2013) proposed a wave equation FWI algorithm:

$$\min_{\mathcal{M},\mathcal{X}} \left\| \mathcal{A}\mathcal{X} - \mathcal{D} \right\|_2^2, \quad s.t. \quad \mathcal{C}(\mathcal{M})\mathcal{X} = \mathcal{S}, \tag{1}$$

where the $\mathcal{A} \in \mathbb{R}^{\mathcal{M} \times \mathcal{N}}$ is linear observation operator, and the $\mathcal{X} \in \mathbb{R}^{\mathcal{N} \times 1}$ is the model wavefield, $\mathcal{D} \in \mathbb{R}^{\mathcal{M} \times 1}$ is observed data; further, in constraint function, $\mathcal{C}(\mathcal{M}) \in \mathbb{R}^{\mathcal{N} \times \mathcal{N}}$ is discretized partial differential equation (PDE) in constrain function, $\mathcal{M} \in \mathbb{R}^{\mathcal{N} \times 1}$ are model parameters, the $\mathcal{S} \in \mathbb{R}^{\mathcal{N} \times 1}$ is source term; $\mathbb{R}$ is regularization function (Aghamiry et al., 2019a); and $\left\| \cdot \right\|_2$ represents the Euclidean norm.

Our research aims to apply a new regularization algorithm to achieve higher accuracy inversions in the presence of high background noise. As mentioned before, we need to combine $\ell_2$ norm with sparsity to make the function capable of sparse prediction. At the same time, we can obtain a tighter convex relaxation by keeping the convexity of the $\ell_1$ norm:

$$\mathrm{conv}(\mathcal{G}_{\mathcal{K}}) \subseteq \left\{ P \,\middle|\, \left\| P \right\|_1 \le \sqrt{\mathcal{K}}, \left\| P \right\|_2 \le 1 \right\} \subsetneq \left\{ P \,\middle|\, \left\| P \right\|_1 \le \sqrt{\mathcal{K}} \right\}, \tag{2}$$

where $\mathcal{K} > 1$ is an adjustable parameter, $P$ the sparse loss data, and $\mathrm{conv}(\mathcal{G}_{\mathcal{K}})$ convex relaxation constrained for loss function $\mathcal{G}_{\mathcal{K}}$ (Belilovsky et al., 2015a). The kernel parameter $\mathcal{K} \in \left\{ 1, \ldots, \mathcal{T} \right\}$ drives the type of the new constraints to behave in a flexible form.

Equation 2 considers the combination with $\ell_2$ regularization on the basis of sparsity and thus can be seen as a better form of a convex constraint than the conventional sparse constraints. The optimization process in traditional FWI also typically uses sparsity to reduce model complexity, while regularization acts as a balance between model complexity and fitting power. In FWI, the optimization process aims to find the optimal medium model that produces synthetic data that best matches the observed data. Since the medium usually has a complex structure and variability, the numerical model needs to contain a large number of parameters to describe it accurately. The more complex the medium or, the noisier the data, can cause the higher the complexity of the model.

The complexity of the model usually directly affects the accuracy and stability of the inversion results. Therefore, in order to obtain better inversion results, a trade-off between model complexity and inversion accuracy needs to be made, and the model complexity needs to be controlled. In this case, regularization methods can be used to limit the model complexity to prevent model overfitting and thus improve the generalization performance of the model. However, using sparsity alone to reduce model complexity may not be sufficient, as sparsity may not capture all possible constraint relationships. In contrast, the use of a combined form of different constraints allows for more careful consideration of the model properties and, thus, a more reasonable reduction of the model complexity for better generalization performance:

$$\left\| \mathcal{M} \right\|_{\mathcal{K}}^{sp} = (\sum_{i=1}^{\mathcal{K}-\mathcal{T}-1} (\left| \mathcal{M} \right|_i^{\downarrow})^2 + \frac{1}{\mathcal{T}+1} (\sum_{i=\mathcal{K}-\mathcal{T}}^{\mathcal{T}} \left| \mathcal{M} \right|_i^{\downarrow})^2)^{\frac{1}{2}}, \tag{3}$$

where the $\left| \mathcal{M} \right|_i^{\downarrow}$ means the $i$th maximum element of the vector in the model parameter matrix $\left| \mathcal{M} \right|$, the $\left\| \ \right\|_{\mathcal{K}}^{sp}$ represents K-support norm, where $\mathcal{K}$ controls the sparsity of the array, and $\mathcal{T}$ is the size of the data (Belilovsky et al., 2015b). When the $\mathcal{T} \in \left\{ 0, 1, ..., \mathcal{K}-1 \right\}$, we can get:

$$\left| \mathcal{M} \right|_{\mathcal{K}-\mathcal{T}-1}^{\downarrow} > \frac{1}{\mathcal{T}+1} \sum_{i=\mathcal{K}-\mathcal{T}}^{\mathcal{T}} \left| \mathcal{M} \right|_i^{\downarrow} \geq \left| \mathcal{M} \right|_{\mathcal{K}-\mathcal{T}}^{\downarrow}. \tag{4}$$

The K-support norm is divided into two components (Fleet et al., 2014). One significant component is the $\ell_2$ norm, and the other small component is the $\ell_1$ norm. The $\kappa$ is an adjustable parameter to balance the proportion of $\ell_1$ norm and $\ell_2$ norm. Therefore, the new norm can provide an adequate balance between the stability and sparsity of the algorithm.

The objective function of FWI based on wavefield reconstruction inversion has reduced the nonlinearity as much as possible by distributed iteration. The wavefield reconstruction inversion algorithm uses regularization terms and model constraints in its optimization to reduce the effect of noise and improve the stability of the inversion. In the form of the wavefield reconstruction inversion algorithm, after iterative optimization, new fitting data, forward results, and adjoint matrices will be obtained concurrently, which can avoid solving the wave equation in the traditional FWI optimization process in a straightforward way (Aghamiry et al., 2021). The wavefield reconstruction inversion algorithm has been an excellent optimization algorithm in recent years, and the K-support norm FWI algorithm is based on its ideas. However, the focus of wavefield reconstruction inversion is on how to avoid local minima. It changes the optimization idea of conventional algorithms to reduce the solution space range by adding constraint conditions, such as iteratively refined-wavefield reconstruction inversion. However, the wavefield reconstruction inversion-like algorithm does not discuss the choice of norm. The K-support norm FWI algorithm focuses more on proposing a new idea of regularization and testing its possibility and prospect of being applied in FWI.

Moreover, the wavefield reconstruction inversion allows the discrete Helmholtz operator to affect the model parameters directly. It reduces the possibility of the objective function tending to be trapped by a local minimum. However, when there is high background noise in the observed data, the objective function of the double penalty term based on wavefield reconstruction inversion cannot effectively deal with the inversion difficulties caused by outliers and noise. Therefore, we propose optimizing the objective function in the form of tighter constraints:

$$\min_{\mathcal{X},\mathcal{M}} \mathcal{R}(\mathcal{M}) \quad s.t. \quad \mathcal{A}\mathcal{X} = \mathcal{D}, \quad \mathcal{C}(\mathcal{M})\mathcal{X} = \mathcal{S}, \tag{5}$$

$$\mathcal{R}(\mathcal{M}) = \beta \|\mathcal{M}\|_{\mathcal{K}}^{sp}, \tag{6}$$

where the $\beta$ is the parameter of the regularization. The value of the $\beta$ and $\mathcal{K}$ could be chosen by cross-validation according to the theory of regularization-based methods (Hastie et al., 2001). In the K-support norm constraint form, we use the $\ell_2$ norm to maintain the stability of the newly added penalty item and the $\ell_1$ norm to weaken the influence of anomalous velocity information in solution space on inversion. In the data processing, we can provide larger weights to model parameters that represent geological information in the optimization process of model parameters; and give smaller weights to insignificant variables, such as noise, to weaken the interference and influence of outliers on model parameters in the optimization process and improve the accuracy and stability of inversion. (Lai et al., 2014). In addition, the $\ell_1$ norm is sparse and unstable, and the $\ell_2$ norm has good stability (Xu et al., 2011); therefore, the K-support norm, which combines the $\ell_1$ and the $\ell_2$ norms, can balance the sparsity and stability of the model.

## ALTERNATING DIRECTION METHOD OF MULTIPLIERS (ADMM)

The ADMM algorithm solves convex optimization problems by decomposing the problem into smaller pieces (Gambella and Simonetto, 2020). This algorithm can be used when the independent variable is composed of two parts and decompose the optimization.

In the optimization part of frequency domain FWI, we continue to use the form of equations 5

and 6 as the objective function; in addition, we adopt ADMM as a method to iteratively solve the newly constructed multivariate optimization problem (Aghazade et al., 2022):

$$\mathcal{X}^{k+1} = \arg\min_{\mathcal{X}} \left( \left\| \mathcal{A}\mathcal{X} - \mathcal{D}^k - \mathcal{D} \right\|_2^2 + \lambda \left\| \mathcal{C}(\mathcal{M}^k)\mathcal{X} - \mathcal{S}^k - \mathcal{S} \right\|_2^2 \right), \tag{7}$$

$$\mathcal{M}^{k+1} = \arg\min_{\mathcal{M}} \left( \mathcal{R}(\mathcal{M}) + \lambda \left\| \mathcal{C}(\mathcal{M})\mathcal{X} - \mathcal{S}^k - \mathcal{S} \right\|_2^2 \right), \tag{8}$$

$$\mathcal{S}^{k+1} = \mathcal{S}^k + \mathcal{S} - \mathcal{C}(\mathcal{M}^{k+1})\mathcal{X}^{k+1}, \tag{9}$$

$$\mathcal{D}^{k+1} = \mathcal{D}^k + \mathcal{D} - \mathcal{A}\mathcal{X}^{k+1}, \tag{10}$$

where $\lambda$ is the penalty parameter, and $\mathcal{C}(\mathcal{M}) \in \mathbb{R}^{\mathcal{N} \times \mathcal{N}}$ is discretized partial differential equation (PDE) in constrain function, $\mathcal{S} \in \mathbb{R}^{\mathcal{N} \times 1}$ is the source term, the $\mathcal{X} \in \mathbb{R}^{\mathcal{N} \times 1}$ is the model wavefield, $\mathcal{D} \in \mathbb{R}^{\mathcal{M} \times 1}$ is observed data, $\mathcal{M} \in \mathbb{R}^{\mathcal{N} \times 1}$ are model parameters. With the iterative update of the objective function, the error caused by the operator selection in the penalty is gradually corrected so that both the minimum of the wavefield and model parameters meet acceptable accuracy at the convergence point (Aghamiry et al., 2022).

## NUMERICAL EXAMPLES

We use three different velocity models to test the performance of FWI based on the K-support norm, namely the Marmousi II, the 2004 BP and SEG/EAGE Overthrust models in this section. We take the frequency domain forward algorithm and add the perfect-matched layer (PML) as the absorbing boundary condition (Pratt, 1999). Moreover, a fixed operator is used in the objective function. With the increase in the number of iterations, when this fixed operator that we adapted takes a reasonable value, the convergence accuracy of the objective function is accurate enough (Aghamiry et al., 2019a). Once we have determined the regularization parameters, we do not iterate over them to update them.

To test the convergence speed of the different algorithms in the model, we adapt the root mean squared error (RMSe) to test the deviation of the fitted model from the true model under different algorithms:

$$\text{RMSe} = \sqrt{\frac{1}{\mathcal{N}_x \times \mathcal{N}_z} \sum_{i=1}^{\mathcal{N}_x \times \mathcal{N}_z} (\frac{\mathcal{M}_{true} - \mathcal{M}_{inv}}{\mathcal{M}_{true}})_i^2}, \tag{11}$$

where the $\mathcal{N}_x$ and $\mathcal{N}_z$ represent the sample scale of the matrix of the true and simulation velocity models (Warner and Guasch, 2016).

In addition, FWI is very susceptible to noise interference. To quantify the effect of noise on inversion results and to compare the denoising ability of different algorithms, we express the level of noise in a signal according to the formula for signal-to-noise ratio (SNR) (de Ridder and Dellinger, 2011):

$$\text{SNR} = 20 * \text{Log}_{10}(\frac{\|\mathcal{D}\|_2}{\|\mathcal{E}\|_2}), \tag{12}$$

where the $\mathcal{E}$ is the noise data. To further demonstrate the suppression effect of the K-support norm on outliers, this paper adopts a low SNR signal of 4.5 $dB$ in three different models.

The combination of $\ell_1$ and $\ell_2$ norms is applied in the objective function to solve the interference of outliers to inversion results. As shown in Figure 1, the quadratic growth rate of the $\ell_2$ norm error is faster, amplifying the negative effects of large noise. However, the linear growth rate of the $\ell_1$ norm is slow, making it insensitive to large noises, thus acting as an implicit suppression of adverse effects, so it is relatively robust. The Huber norm behaves as the $\ell_2$ norm when it is less than the threshold and the $\ell_1$ norm when it is greater than the threshold.

In contrast, the K-support norm is a norm that strikes a balance between sparsity and smoothness. It is a regularization where the parameter $\kappa$ controls the balance between sparsity and smoothness.

As shown in Figures 2 (A) and (E), the contour line of the $\ell_1$ norm shows sparsity due to non-differentiability at zero. The contour line of the $\ell_2$ norm shown in Figure 2 (B) and (F) is a circle, indicating that it is likely to select more number of eigenvalues. Figures 2 (C) and (J) are the visual explanation and contour graph of the Huber norm. The most basic K-support norm form can be obtained, as shown in Figure 2 (D) and (H). Unlike the $\ell_1$ norm only biases sparsity, the $\ell_2$ norm does not bias sparsity. On the other hand, the K-support norm is non-differentiable when the $\kappa$ value equals the unity (The K-support norm is completely backward to $\ell_1$ norm); in such cases, the K-support exhibits the form of a thorough $\ell_1$ norm. However, this new norm form can produce sparsity or is differentiable when the $\kappa$ value is larger than the unity; therefore, the new norm form prefers to produce reasonable sparsity compared with the $\ell_1$ norm.

Figures 1 and 2 show that the Huber norm is smooth in the region near zero and linear in the area far from zero. The nature of the Huber norm better fits the data in the presence of outliers, as it is robust to them. However, the K-support norm retains varying degrees of regularization properties and has a shape between the diamond shape of the $\ell_1$ norm and the circle shape of the $\ell_2$ norm. The size of the K-support norm is controlled by the parameter $\kappa$; the larger it becomes, the larger the value of the K-support norm is, and vice versa. It is not a simple combination of the $\ell_1$ and the $\ell_2$ norms, but the constraint is adjusted by the $\kappa$ parameter. When we do not need this combination, it can also present the characteristics of the $\ell_1$ norm or the $\ell_2$ norm completely by adjusting the parameter $\kappa$. Therefore, in practical applications, a larger value of the K-support norm can be used as the regularization term when the predictive power of the model needs to be improved. In comparison, a smaller value of the K-support norm can be used when the explanatory power of the model needs to be improved, or the complexity of the model needs to be reduced.

Thus the K-support norm encourages sparsity of the weights, which can produce more sparse solutions, reducing the complexity of the model and improving its interpretation. And the most important is that it allows more flexibility to adjust the parameter $\kappa$ to achieve different regularization effects, thus improving the prediction performance of the model.

**Marmousi II Model**

At first, we test the performance of the K-support norm in the Marmousi II model. Our model is 17.0 km long and 3.5 km deep. The observation system consists of 140 equidistant sources. We adopt the multi-scale inversion strategy in the frequency domain, which ranges from 1 Hz to 8 Hz. An equal-width PML boundary surrounds the model. In the forward modeling process, we use the conventional one-dimensional linearly increasing velocity model as the initial model, in which the velocity ranges linearly from 1km / s to 4.25 km/s.

To test the effect of K-support in the deep part of the low SNR model, we do not use the clipped Marmousi II. Figure 3 (A) is the true velocity model, and Figure 3 (B) is the initial velocity model. Figure 4 (A) and Figure 4 (B) show the Tikhonov regularization FWI inversion results and the FWI inversion results based on the K-support norm, respectively, after adding 4.5 *dB* noise. Figures 4 (C) and Figure 4 (D) are the difference between the results of the Tikhonov regularization FWI and the true velocity model, and the difference between the results of the K-support norm FWI and the true velocity model, respectively. In Figure 5 (A), the solid red line is the convergence speed based on the K-support norm, and the blue line is the convergence speed of Tikhonov regularization FWI. The new algorithm shows better convergence speed owing to the ADMM. In Figure 5 (B), we use the RMS error in equation 11 to reflect the effects of modified FWI. To pay more attention to the details, we choose six different $X$ positions for comparing velocity profiles. As shown in Figure 6, there are six positions $X = 8.72$ km, $X = 9.60$ km, $X = 10.92$ km, $X = 12.56$ km, $X = 13.40$ km, $X = 15.28$ km, respectively. These six positions are evenly distributed near the layered structure in the middle of the model. The solid red line in the figure represents the inversion results of the K-support norm FWI. Compared with the Tikhonov regularization FWI based on the $\ell_2$ norm, the new algorithm has a better suppression effect on noise, which reflects in the peak value of the velocity model, and the fitting in the use of the K-support norm FWI is better. To better reflect the noise suppression effect of the algorithm, we adopt a more challenging form of velocity analysis. We select four slices at different depths, i.e., $Y = 2.24$ km, $Y = 2.40$ km, $Y = 2.72$ km, and $Y = 3.00$ km, respectively, as shown in Figure 7. Comparing the four vertical velocity profiles at different depths shows that the K-support norm has pronounced suppression effects on noise since there are no violent saw-tooth between velocity peaks. The above results show that the sparsity of the $\ell_1$ norm has brought great help to the improvement of the inversion result.

**2004 BP Model**

To better explain the applicability of the proposed algorithm, we further tested the middle part of the 2004 BP model. The middle part of the model describes the geological characteristics of the Eastern / Central Gulf of Mexico and offshore Angola. Because the middle part of the model is composed of a high-velocity salt body, the division of the salt body is one of the difficulties in the inversion of the model. In addition, the channels are located near the salt body, which will further affect the velocity inversion and increase the difficulty of inversion (Billette and Brandsberg-Dahl, 2005). Figure 8 (A) is the actual velocity model in the middle of the 2004 BP model, and Figure 8 (B) is the initial velocity model, with a linear velocity increase ranging from 1.5 km / s to 5 km / s. To test the effect of noise, we add random background noise to the recorded data set, as shown in Figure 9 (A), more detailed, as shown in Figure 9 (B). The SNR of the noise-added data set is 4.5 *dB*. Figure 9 (C) shows the added noise. In the case of 4.5 dB background noise, the inversion results based on Tikhonov regularization FWI are depicted in Figure 10 (A). With increasing the number of iterations, low SNR and outliers will significantly impact the inversion result, and the wrong velocity structure is created, especially in places with significant velocity changes. However, FWI, based on the K-support norm, can avoid this situation as much as possible. Figure 10 (B) shows that FWI based on the K-support norm can still obtain roughly accurate inversion results even in the case of low SNR, not only in the channels but also in the inversion results of the salt body. Figure 10 (C) and Figure 10 (D) are the difference between the results of the Tikhonov regularization FWI and the true velocity model, and the difference between the results of the K-support norm FWI and the true velocity model, respectively. Further, we use multi-scale frequency division inversion and obtain inversion results at each frequency from 1.04 Hz to 5.35 Hz in Figure 11. It can be seen that the FWI based on the K-support norm can avoid the interference of outliers to a certain extent and can obtain correct inversion results faster than the Tikhonov regularization FWI, especially for salt bodies and channels more accurately. For the performance of K-support under the 2004 BP model, we quantitatively evaluate it in two ways. The first is the comparison of convergence speed, as shown in Figure 12 (A). The solid red line is the convergence speed based on the K-support norm, while the blue from the Tikhonov regularization FWI. With the increase in iterations, the convergence speed of the proposed algorithm is better. Secondly, we compare the convergence of model RMS error defined by equation 11, as shown in Figure 12 (B). With the increasing number of iterations, the model RMS error based on the K-support norm FWI will be much smaller than the Tikhonov regularization FWI when a low SNR signal is used. The results show that the model RMS error value does not converge all the time but will remain near a not unacceptable error range, increasing the inversion deviation caused by noise and outliers. In contrast, the FWI based on the K-support norm performs well under low SNR, and there is no increase in model RMS error, which proves that it has an apparent inhibitory effect on noise and outliers. We also compare the velocity profiles at six different $X$ positions of the inversion results. Figures 13 (A-F) show the vertical velocity profiles at six positions $X = 7.4$ km, $X = 7.8$ km, $X = 8.2$ km, $X = 9.8$ km, $X = 10.4$ km and $X = 11.3$ km, respectively, while Figures 14 (A-F) show the horizontal velocity profiles at six positions $Y=2.83$ km, $Y=3.35$ km, $Y=3.91$ km, $Y=4.47$ km, $Y=4.61$ km, $Y=5.03$ km. It can be seen from the results that the oscillatory solution of FWI based on the K-support norm is not evident, which proves that it has a good suppression effect on noise. Especially in some high-speed areas, it can peak close to the true velocity, and there will be no undeniable velocity change in low-velocity areas.

**SEG/EAGE Overthrust Model**

We further tested the performance of the modified method in the SEG/EAGE Overthrust Model. The SEG/EAGE Overthrust Model is a commonly used geological model in seismic exploration and is used primarily to describe geological structures with overthrust structures. Figure 15(A) shows the actual velocity model of the SEG/EAGE Overthrust Model, and Figure 15(B) indicates the initial velocity model. We used the smoothed data as the initial model. Subsequently, we add a random noise field to the seismic wave field. As shown in Figure 16 (a) for the noise-free data, Figure 16 (b) for the random noise of 4.5 *dB*, and Figure 16 (c) for the dataset with added noise. The results of the FWI based on two different regularization algorithms are shown in Figures 17 (a-j), where (a-e) are the Tikhonov regularization FWI inversion results in 3.22 Hz, 5.71 Hz, 7.59 Hz, 9.19 Hz, and 11.12 Hz, respectively, and (f-j) are the FWI based on K-support norm inversion results in 3.22 Hz, 5.71 Hz, 7.59 Hz, 9.19 Hz, and 11.12 Hz, respectively. Figures 17 (k-o) are the differences between Tikhonov FWI and the true velocity model. Figures 17 (p-t) are the differences between K-support FWI and the true velocity model. The random noise, whose information is shown in Figure 16, was added in these cases. In order to compare the inversion effect of the two algorithms more accurately, we conducted the comparison at six positions in the horizontal direction and six positions in the vertical direction, respectively. The results of the comparison in six of the horizontal directions are shown in Figure 18, and we have marked the places with relatively significant differences with blue arrows. The black dashed line in the figure is the initial velocity, the solid black line is the true velocity, the solid blue line is the FWI based on the Tikhonov norm, and the solid red line is the K-support norm FWI. Similarly, the comparisons for the six vertical directions are shown in Figure 19.

In addition, we evaluate the proposed algorithm and the Tikhonov algorithm in terms of the suppression of other types of noises besides random noise. Figure 20 (a) shows the real part of the clean wavefield, Figure 20 (b) shows the hyperbolic noise, and Figure 20 (c) shows the real part after adding the noise to the wavefield. We compared the two algorithms separately, using the multi-scale frequency domain algorithm, for four different sets of frequencies, as shown in Figure 21. Figures 21 (a-d) are the inversion results based on Tikhonov regularization, while (e-

h) are the inversion results based on the proposed algorithm. The results show that the inversion effect based on the proposed algorithm is better, especially in the visible description of the layer structure. The distinction between high and low velocity layers at the bottom becomes more evident, which verifies the suppression effect of the improved algorithm under different types of noise conditions.

## DISCUSSIONS

The non-linear least-squares problem is solved in the FWI. Sometimes, it becomes difficult to achieve convergence at the optimal global solution in the model space. The wavefield reconstruction inversion method was proposed for solving this non-linear optimization problem. Simultaneously, the various sources of wavefield reconstruction inversion are independent and uncorrelated from each other, so the nonlinearity of the algorithm is significantly reduced (van Leeuwen and Herrmann, 2016). However, the optimization process of the objective function consumes much computational cost due to the large size of the data. Thus, the first algorithm that should be considered from a theoretical point of view is the Dual Ascent (Wong, 1984). However, the dual ascent has strict requirements for functions in practical application, and non-strict convex functions are not applicable. An improved method, the alternating direction multiplier method (ADMM), is proposed to integrate the decomposability and the excellent convergence properties of the dual ascent and the multipliers methods. The ADMM algorithm is mainly used in the case of an enormous solution space. Its advantage is to decouple the problem so that it can be distributed and parallel computing (Boyd et al., 2011). At the same time, because the idea of the alternating solution of ADMM is very consistent with the combination of two mutually independent norms in the K-support algorithm, we use K-support based on the ADMM solution to replace the conventional misfit function in FWI.

Here, we describe the possibility and necessity of improving the objective function and the possible problems. FWI, as a fitting data algorithm, actually relies more on understanding optimization from the mathematics point of view. Not only the FWI of the conventional process, even the FWI based on machine learning and artificial intelligence, which will become a hot research topic in the future, can be transformed into an optimization problem in its essence. The optimization problem itself can be divided into two sub-problems: the first one is how to construct a proper misfit function, and the second one is how to optimize this misfit function. Therefore, how to construct and optimize an appropriate objective function should become the focus of our research.

The regularization can limit the complexity of the model to a certain extent, enhance the normalization ability of the model, and avoid overfitting. At the same time, when solving the optimization problem directly, the norm is added, which can effectively deal with the problems of numerical instability and slow convergence caused by the irreversibility of the data matrix. In addition, the polynomial objective function is dominated by a quadratic regular term, which makes the convexity of the objective function at least the level of a quadratic function.

One example is the K-support norm. We have made corresponding attempts in this paper. The numerical experimental results of the Marmousi II model and the 2004 BP model align with expectations. When the data contains high background noise and outliers, their effects on inversion results based on the new adjustable objective function form with sparsity are significantly suppressed than the conventional algorithm.

Another important issue is the tuning parameter of the algorithm. In the Lagrangian algorithm, the value of the Lagrangian operator $\lambda$ is critical. In fact, in FWI, the optimal value of $\lambda$ differs according to different models. However, some literature has proved that as long as the value of $\lambda$ is reasonable, the convergence accuracy of the objective function is accurate enough (Aghamiry et al., 2019a). However, if the same $\lambda$ value is used for different models, the inversion results could be affected significantly. This phenomenon is the same as other polynomial constraint forms. The selection of parameters is not invariable but should be adjusted according to different models. Therefore, this problem cannot be ignored in numerical simulation. The simplest solution is to carry out multiple inversions by adjusting parameters and selecting a reasonable parameter by balancing inversion accuracy and time consumption. A more complex algorithm changes the parameters into an adaptive form. The adaptive algorithm associates the operator with the independent variables in the polynomial to obtain a parameter that can be self-corrected with the model. This algorithm is in line with our idea in theory, but there are not many numerical simulations to prove its effectiveness. Therefore, in the future, for FWI, especially for the optimization part, the more important direction is to find more appropriate and rigorous constraint forms. This direction is crucial for conventional FWI and artificial intelligence FWI, which will be more mainstream in the future.

## CONCLUSIONS

For the problem that conventional FWI is sensitive to non-Gaussian white noise and outliers, this paper proposes a new algorithm to suppress their influence on FWI results. Based on the wavefield reconstruction inversion, this paper first attempts to linearize the constraint form and then adopts a new form of combining multiple norms to strengthen the robustness and generalization ability of the objective function by adjusting tuning parameters. The ADMM algorithm reduces the computational complexity from polynomial to linear in the objective function's iterative process to speed up the convergence rate. In the numerical experiments, we compare the inversion results of our K-support norm with those of conventional FWI schemes using the Marmousi II model and 2004 BP model under the same low SNR condition. The results show that the FWI based on the K-support norm has better convergence and noise suppression effects. Under the condition of low SNR with 4.5 dB noise in the observation data, more accurate inversion results can be obtained at the bottom of the Marmousi II model and the salt body part of the 2004 BP model by mitigating high sensitivity to noise and outliers in conventional FWI. Therefore, our proposed method can realize high-resolution imaging of deep regions of complex structures.

## APPENDIX A. WAVEFIELD-RECONSTRUCTION INVERSION (WRI)

van Leeuwen and Herrmann proposed a wave equation FWI algorithm based on the penalty function.

The regularization term of PDE constrained form can be expressed as follows:

$$\min_{\mathcal{X},\mathcal{M}} \left\| \mathcal{A}\mathcal{X} - \mathcal{D} \right\|_2^2, \quad s.t. \quad \mathcal{C}(\mathcal{M})\mathcal{X} = \mathcal{S}, \tag{A-1}$$

where $\left\| \bullet \right\|_2^2$ is the $\ell_2$ norm, and the $\mathcal{M} \in \mathbb{R}^{N \times 1}$ is the model parameters; $\mathbb{R}$ is the regularization function, contains preliminary information about model parameters, $\mathcal{X} \in \mathbb{R}^{N \times 1}$ represents the wavefield, $\mathcal{D} \in \mathbb{R}^{M \times 1}$ is the recorded seismic data, $\mathcal{S} \in \mathbb{R}^{N \times 1}$ is the source term, and the linear observation operator $\mathcal{A} \in \mathbb{R}^{M \times N}$ sampling $\mathcal{X}$ at the receiver positions.

From the mathematical point of view, the most appropriate way to solve this type of constrained optimization problem is in the form of a Lagrangian function:

$$\min_{\mathcal{M},\mathcal{X}} \max_{\mathcal{V}} \mathcal{F}(\mathcal{M},\mathcal{X},\mathcal{V}) = \min_{\mathcal{M},\mathcal{X}} \max_{\mathcal{V}} \left\| \mathcal{A}\mathcal{X} - D \right\|_2^2 + \mathcal{V}^T \left[ \mathcal{C}(\mathcal{M})\mathcal{X} - \mathcal{S} \right], \tag{A-2}$$

where $\mathcal{V} = \left[ \mathcal{V}_1, \mathcal{V}_2, \ldots \right]$. The $\nu$ represents the Lagrange multiplier. The advantage of using the Lagrange function is that after iterative optimization, new fitting data, forward results, and an adjoint matrix will be obtained simultaneously, which avoids the need for an explicit solution in the optimization process of conventional FWI, which means that the wavefield reconstruction inversion algorithm obtains a solution consistent with augmented wave equation term.

Since the model parameter $c$ based on the PDE operator may converge to an approximate minimum value when the start model is not ideal, therefore, van Leeuwen and Herrmann redefine the primitive constrained problem into a quadratic penalty problem in 2013:

$$\min_{\mathcal{M},\mathcal{X}} \mathcal{F}(\mathcal{M},\mathcal{X}) = \min_{\mathcal{M},\mathcal{X}} \left\| A\mathcal{X} - D \right\|_2^2 + \lambda \left\| \mathcal{C}(\mathcal{M})\mathcal{X} - \mathcal{S} \right\|_2^2. \tag{A-3}$$

With the form of a quadratic penalty term, the original full space-constrained form can be turned into a double penalty term form, reducing the complexity of the algorithm while improving the ability of FWI to converge to the more exact minimum value when the initial model is poor.

LIST OF FIGURES

**Figure 1.** Linear growth of the $P$ (=0.1 and 0.5) norm, the $\ell_1$ norm, the $\ell_2$ norm, the Huber norm and the K-support norm. The horizontal and vertical axes are weight $w$ and norm($w$), respectively.

**Figure 2.** Visual explanations and contour graphs. Figures (A-D) and (E-H) are the bird's eye vires of value surfaces and the contour graphs for the $\ell_1$, the $\ell_2$, the Huber, and the K-support norms, respectively. The contour of the $\ell_1$ norm is diamond-shaped; it has spikes that happen to be at sparse points. The $\ell_2$ norm is circular, which does not produce sparsity. The Huber norm is a combination of $\ell_1$ norm and $\ell_2$ norm controlled by a threshold operator. In contrast, the K-support norm retains varying degrees of regularization properties and has a shape between the diamond shape of the $\ell_1$ norm and the circle shape of the $\ell_2$ norm.

**Figure 3.** Marmousi II model. (A) True velocity model. (B) Initial velocity model.

**Figure 4.** Marmousi II model. (A) Inversion result of the Tikhonov regularization FWI. (B) Inversion result of K-support norm FWI. (C) Difference between Figure 4-A and the true benchmark. (D) Difference between Figure 4-B and the true benchmark. The Tikhonov regularization FWI inversion results and the FWI inversion results are based on the K-support norm, respectively, after adding 4.5 $dB$ noise.

**Figure 5.** Model curves for the test. (A) Misfit error, (B) Model RMS error. The solid red line represents the K-support norm, and the blue line represents the Tikhonov regularization FWI. The formula of model RMS error is shown in Equation 11.

**Figure 6.** Marmousi II model. 1-D velocity models at different $X$-positions. (A) $X$=8.72 km; (B)

*X*=9.60 km; (C) *X*=10.92 km; (D) *X*=12.56 km; (E) *X*=13.40 km; (F) *X*=15.28 km. The vertical comparison of the actual velocity model (solid black line), initial velocity model (grey dotted line), the Tikhonov regularization FWI velocity model (solid blue line), and the K-support norm FWI velocity model (solid red line).

**Figure 7.** Marmousi II model. 1-D velocity models at different *Y*-positions. (A) *Y*=2.24 km; (B) *Y*=2.40 km; (C) *Y*=2.72 km; (D) *Y*=3.00 km. The lateral comparison of the actual velocity model (solid black line), initial velocity model (grey dotted line), the Tikhonov regularization FWI velocity model (solid blue line), and the K-support norm FWI velocity model (solid red line).

**Figure 8.** 2004 BP model. (A) True velocity model. (B) Initial velocity model.

**Figure 9.** Observed data with (A) no noise, (B) 4.5 *dB* random noise, and (C) added noise. The SNR of the noise-added data set is 4.5 *dB*, and the formula of SNR is shown in Equation 12. The difference between the two data sets is an additional noise data set, and all panels are illustrated at the same amplitude.

**Figure 10.** 2004 BP model. (A) Inversion result of Tikhonov regularization FWI. (B) Inversion result of the K-support norm FWI. Both were added 4.5 *dB* noise. (C) Difference between Figure 10-A and the true benchmark. (D) Difference between Figure 10-B and the true benchmark. The 4.5 *dB* random noise is shown in Figure 9.

**Figure 11.** 2004 BP model. (a-f) Tikhonov regularization FWI inversion results in 1.04 Hz, 1.79 Hz, 2.15 Hz, 2.58 Hz, 3.72 Hz, and 5.35 Hz, respectively. (g-l) FWI based on K-support norm inversion results in 1.04 Hz, 1.79 Hz, 2.15 Hz, 2.58 Hz, 3.72 Hz, and 5.35 Hz, respectively. (m-r) Differences between the Tikhonov FWI and the true velocity model. (s-x) Differences between the K-support FWI and the true velocity model. It can be seen that sparsity plays an influential

role in suppressing outliers and noise.

**Figure 12.** Model curves for the test. (A) Misfit error, (B) Model RMS error. The solid red line represents the K-support norm, and the blue line represents the Tikhonov regularization FWI. The formula of the Model RMS error is shown in Equation 11.

**Figure 13.** 2004 BP model.1-D velocity models at different *X*-positions. (A) *X*=7.4 km; (B) *X*=7.8 km; (C) *X*=8.2 km; (D) *X*=9.8 km;(E) *X*=10.4 km; (F) *X*=11.3 km. The vertical comparison of the actual velocity model (solid black line), initial velocity model (grey dotted line), the Tikhonov regularization FWI velocity model (solid blue line), and the K-support norm FWI velocity model (solid red line).

**Figure 14.** 2004 BP model. 1-D velocity models at different *Y*-positions. (A) *Y*=2.83 km; (B) *Y*=3.35 km; (C) *Y*=3.91 km; (D) *Y*=4.47 km; (E) *Y*=4.61 km; (F) *Y*=5.03 km. The lateral comparison of the actual velocity model (solid black line), initial velocity model (grey dotted line), the Tikhonov regularization FWI velocity model (solid blue line), and the K-support norm FWI velocity model (solid red line).

**Figure 15.** SEG/EAGE Overthrust Model. (A) True velocity model. (B) Initial velocity model.

**Figure 16.** The real part of the 3.22 Hz data for the SEG/EAGE Overthrust model. (a) Noise-free data; (b) Random noise; (c) Noisy data.

**Figure 17.** SEG/EAGE Overthrust Model. (a-e) Tikhonov regularization FWI inversion results in 3.22 Hz, 5.71 Hz, 7.59 Hz, 9.19 Hz, and 11.12 Hz, respectively. (f-j) FWI based on K-support norm inversion results in 3.22 Hz, 5.71 Hz, 7.59 Hz, 9.19 Hz, and 11.12 Hz, respectively. (k-o) Differences between the Tikhonov FWI and the true velocity model. (p-t) Differences between the K-support FWI and the true velocity model. Both were added to the random noise; the noise

information is shown in Figure 16.

**Figure 18.** SEG/EAGE Overthrust Model, 1-D velocity models at different $Y$-positions. (A) $Y =$ 0.4 km; (B) $Y =$ 0.6 km; (C) $Y =$ 1.2 km; (D) $Y =$ 1.8 km; (E) $Y =$ 2.8 km; (F) $Y =$ 3.2 km. The lateral comparison of the actual velocity model (solid black line), initial velocity model (grey dotted line), the Tikhonov regularization FWI velocity model (solid blue line), and the K-support norm FWI velocity model (solid red line).

**Figure 19.** SEG/EAGE Overthrust Model, 1-D velocity models at different $X$-positions. (A) $X =$ 1.83 km; (B) $X =$ 3.77 km; (C) $X =$ 4.67 km; (D) $X =$ 7.40 km; (E) $X =$ 13.43 km; (F) $X =$ 15.70 km. The vertical comparison of the actual velocity model (solid black line), initial velocity model (grey dotted line), the Tikhonov regularization FWI velocity model (solid blue line), and the K-support norm FWI velocity model (solid red line).

**Figure 20.** The real part of the 3.22 Hz data for the SEG/EAGE Overthrust model. (a) Noise-free data; (b) Hyperbolic noise; (c) Noisy data.

**Figure 21.** SEG/EAGE Overthrust Model. (a-d) Tikhonov regularization FWI inversion results in 3.22 Hz, 6.28 Hz, 7.59 Hz, and 11.12 Hz, respectively. (e-h) FWI based on K-support norm inversion results in 3.22 Hz, 6.28 Hz, 7.59 Hz, and 11.12 Hz, respectively. Both were added to the hyperbolic noise; the noise information is shown in Figure 20.

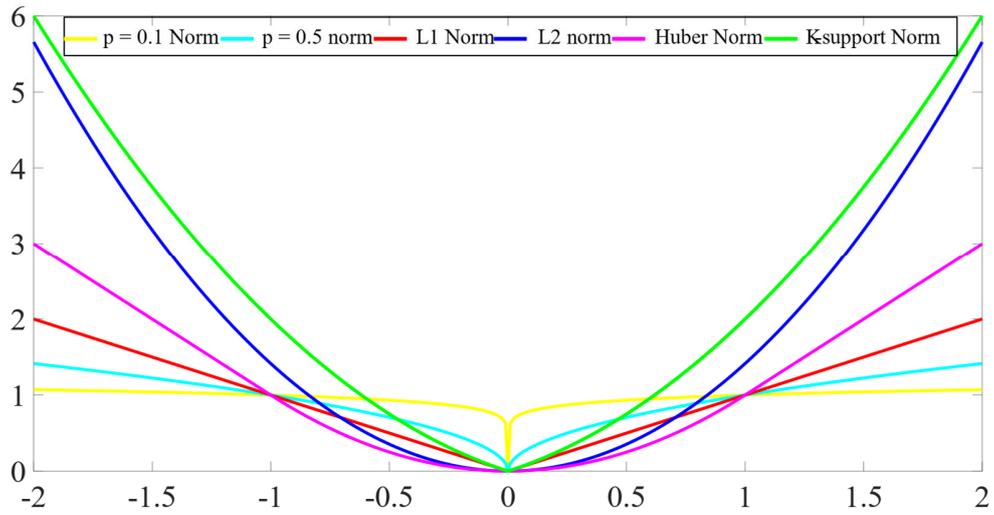

**Figure 1.** Linear growth of the $P$ (=0.1 and 0.5) norm, the $\ell_1$ norm, the $\ell_2$ norm, the Huber norm and the K-support norm. The horizontal and vertical axes are weight $w$ and norm($w$), respectively.

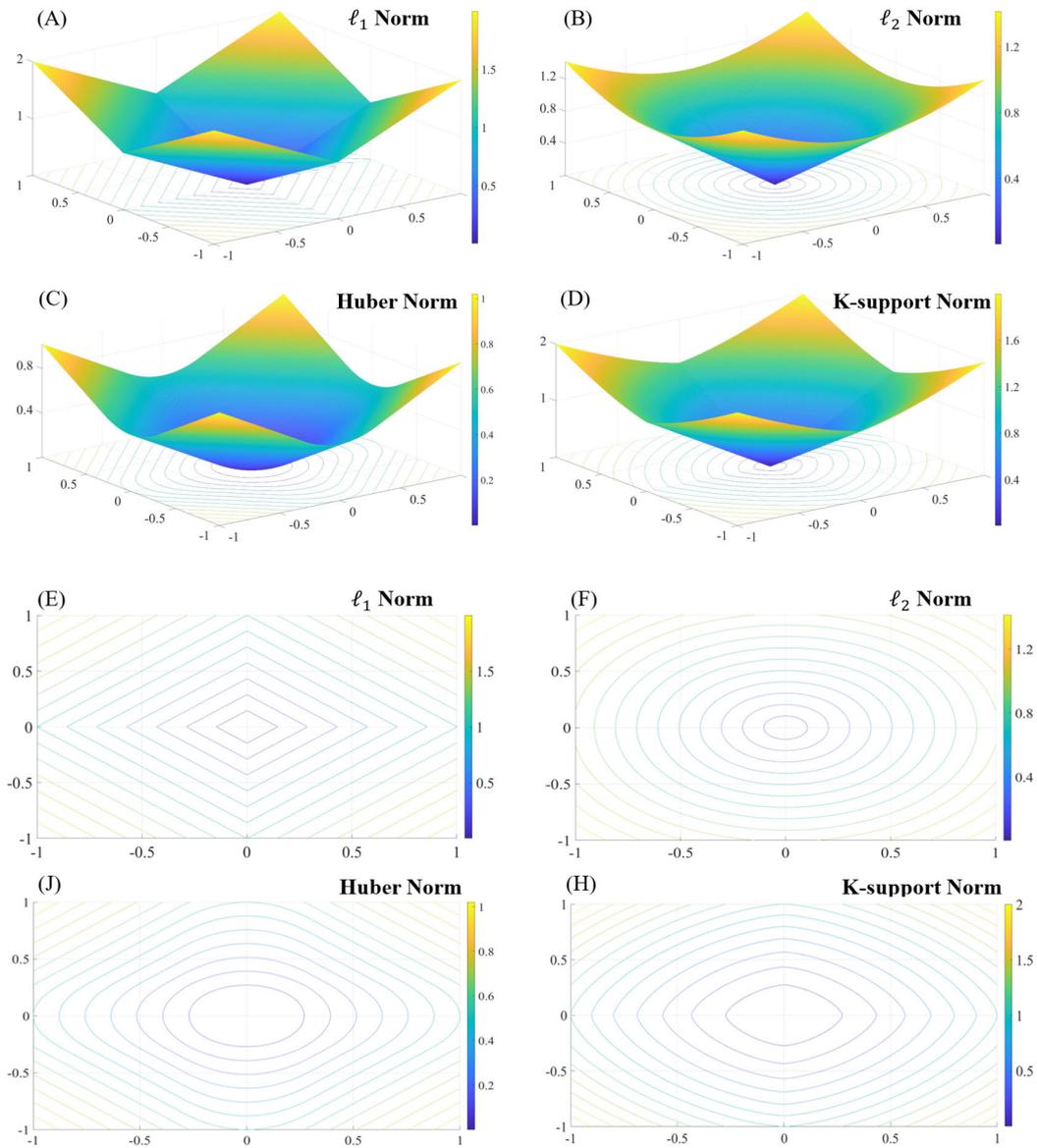

**Figure 2.** Visual explanations and contour graphs. Figures (A-D) and (E-H) are the bird's eye vires of value surfaces and the contour graphs for the $\ell_1$, the $\ell_2$, the Huber, and the K-support norms, respectively. The contour of the $\ell_1$ norm is diamond-shaped; it has spikes that happen to be at sparse points. The $\ell_2$ norm is circular, which does not produce sparsity. The Huber norm is a combination of $\ell_1$ norm and $\ell_2$ norm controlled by a threshold operator. In contrast, the K-support norm retains varying degrees of regularization properties and has a shape between the diamond shape of the $\ell_1$ norm and the circle shape of the $\ell_2$ norm.

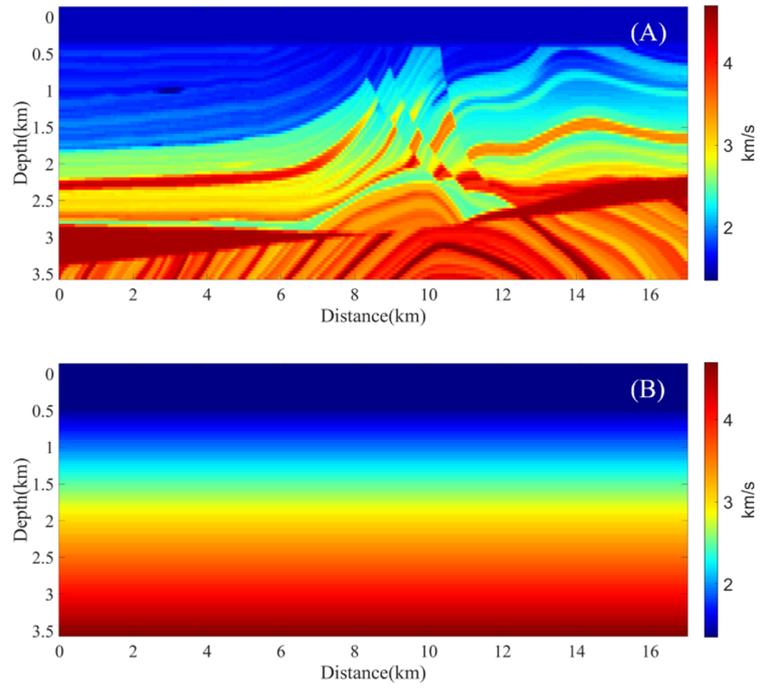

**Figure 3.** Marmousi II model. (A) True velocity model. (B) Initial velocity model.

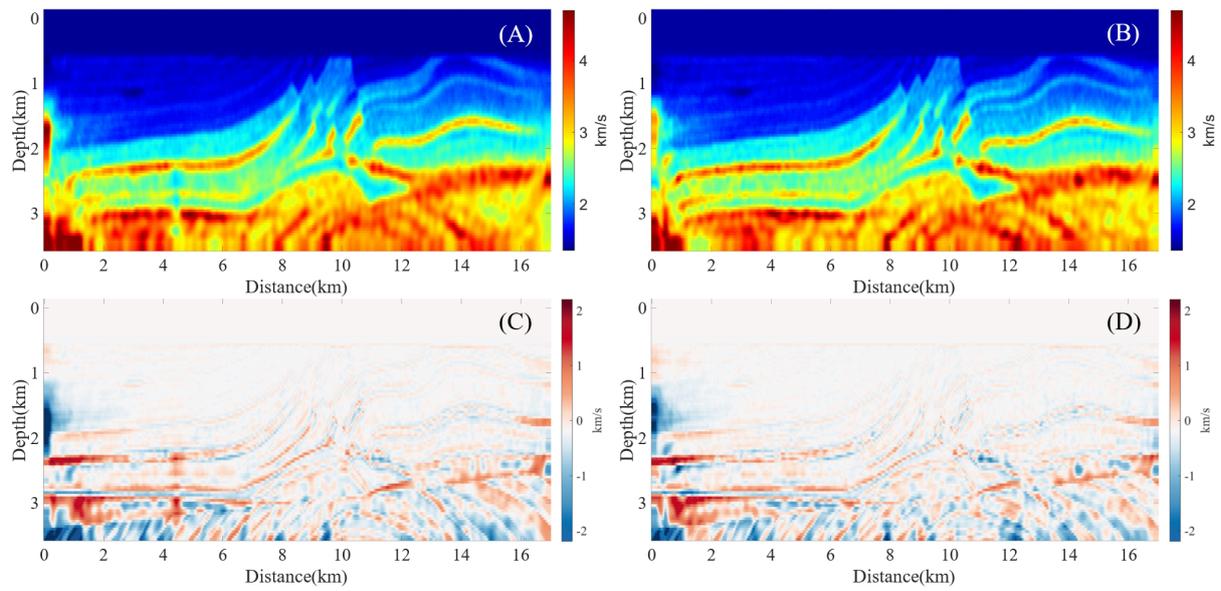

**Figure 4.** Marmousi II model. (A) Inversion result of the Tikhonov regularization FWI. (B) Inversion result of K-support norm FWI. (C) Difference between Figure 4-A and the true benchmark. (D) Difference between Figure 4-B and the true benchmark. The Tikhonov regularization FWI inversion results and the FWI inversion results are based on the K-support norm, respectively, after adding 4.5 *dB* noise.

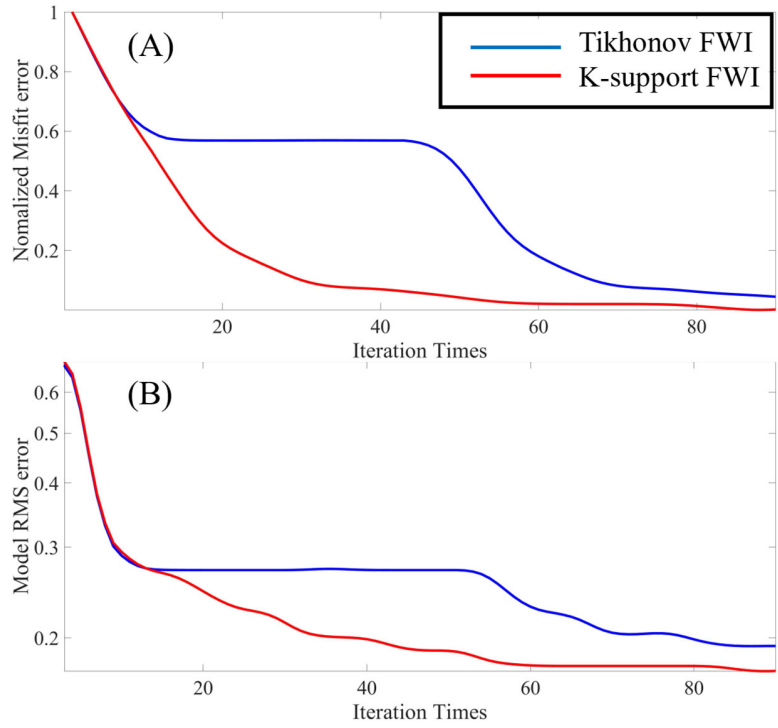

**Figure 5.** Model curves for the test. (A) Misfit error, (B) Model RMS error. The solid red line represents the K-support norm, and the blue line represents the Tikhonov regularization FWI. The formula of model RMS error is shown in Equation 11.

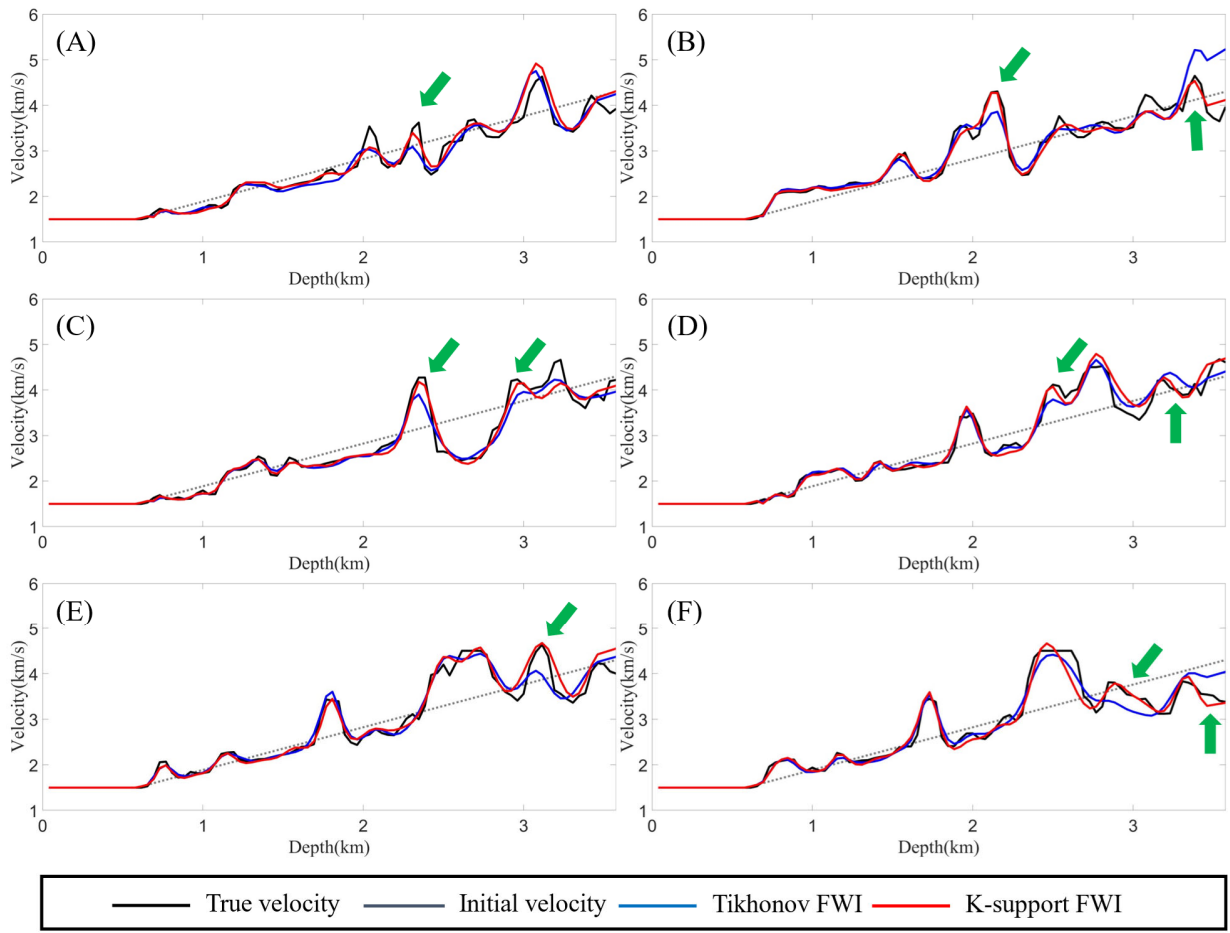

**Figure 6.** Marmousi II model. 1-D velocity models at different *X*-positions. (A) *X*=8.72 km; (B) *X*=9.60 km; (C) *X*=10.92 km; (D) *X*=12.56 km; (E) *X*=13.40 km; (F) *X*=15.28 km. The vertical comparison of the actual velocity model (solid black line), initial velocity model (grey dotted line), the Tikhonov regularization FWI velocity model (solid blue line), and the K-support norm FWI velocity model (solid red line).

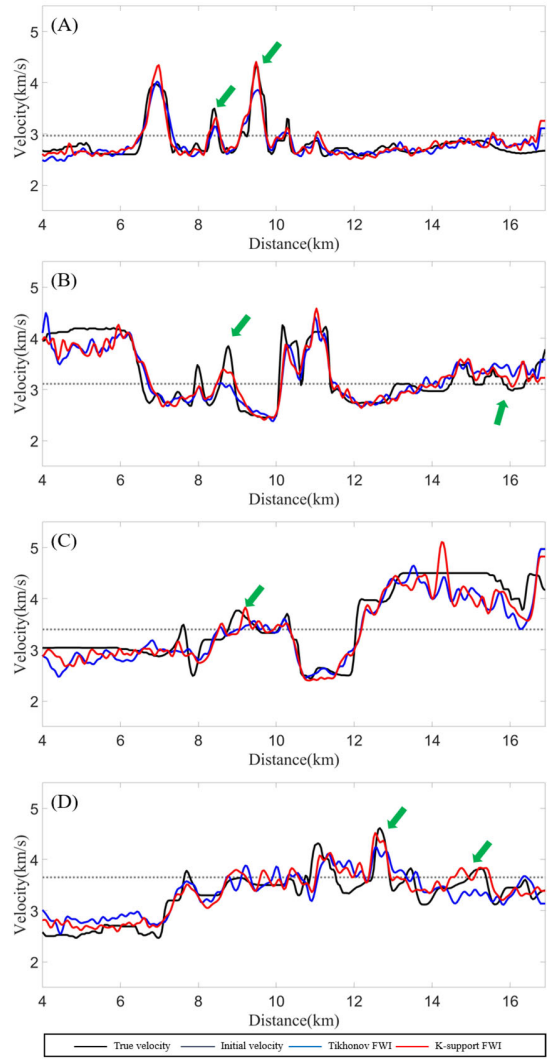

**Figure 7.** Marmousi II model. 1-D velocity models at different *Y*-positions. (A) *Y*=2.24 km; (B) *Y*=2.40 km; (C) *Y*=2.72 km; (D) *Y*=3.00 km. The lateral comparison of the actual velocity model (solid black line), initial velocity model (grey dotted line), the Tikhonov regularization FWI velocity model (solid blue line), and the K-support norm FWI velocity model (solid red line).

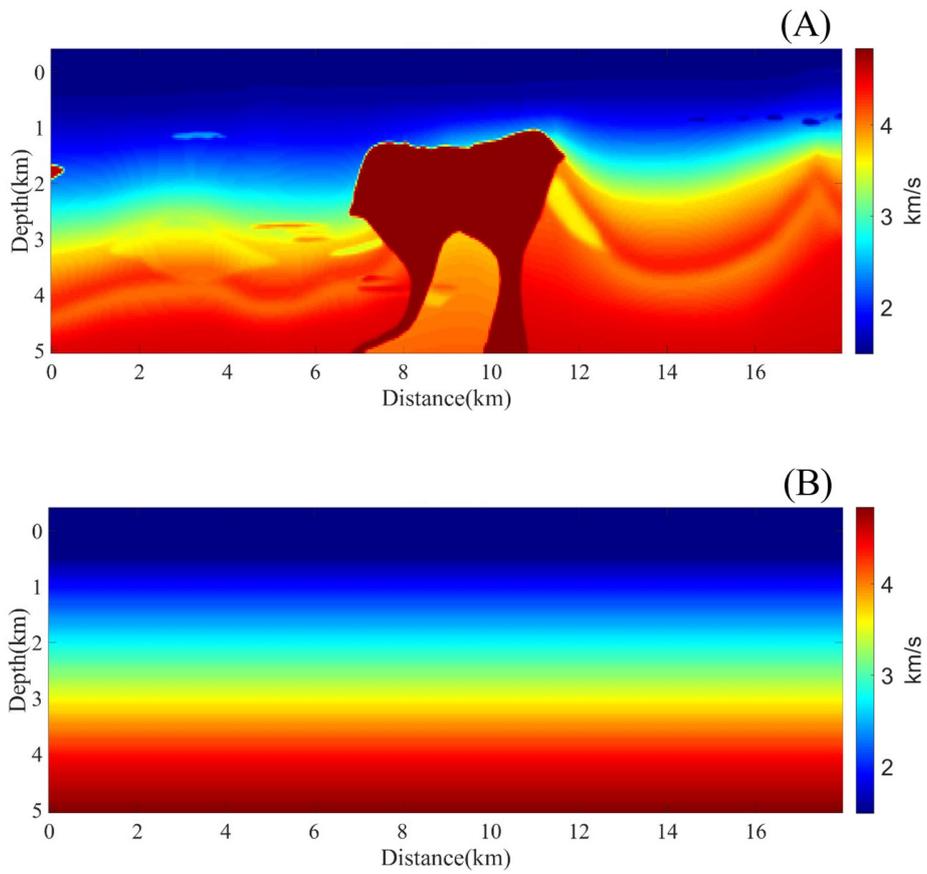

**Figure 8.** 2004 BP model. (A) True velocity model. (B) Initial velocity model.

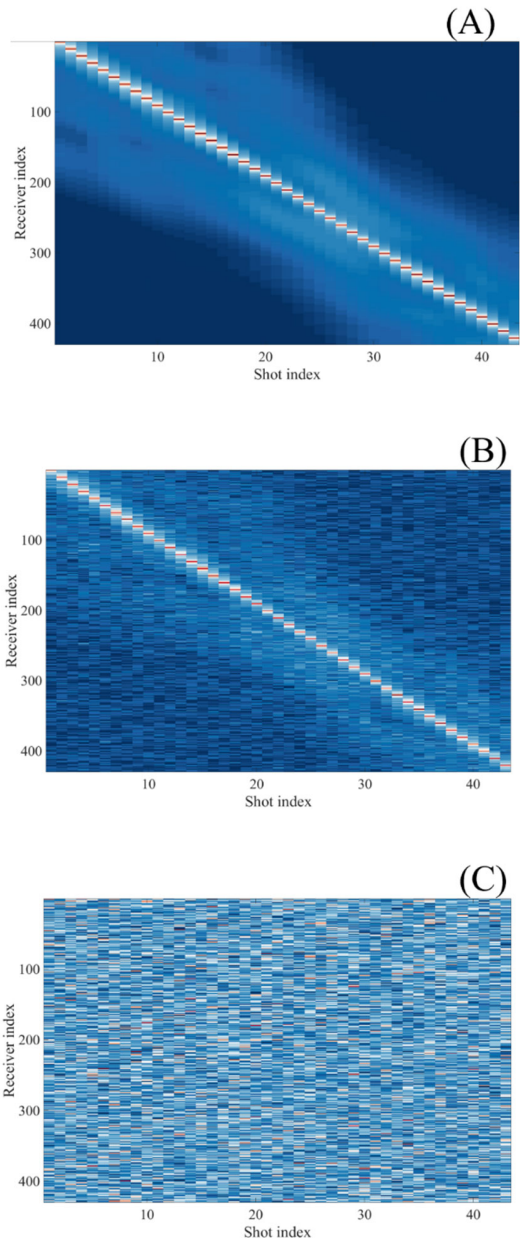

**Figure 9.** Observed data with (A) no noise, (B) 4.5 *dB* random noise, and (C) added noise. The SNR of the noise-added data set is 4.5 *dB*, and the formula of SNR is shown in Equation 12. The difference between the two data sets is an additional noise data set, and all panels are illustrated at the same amplitude.

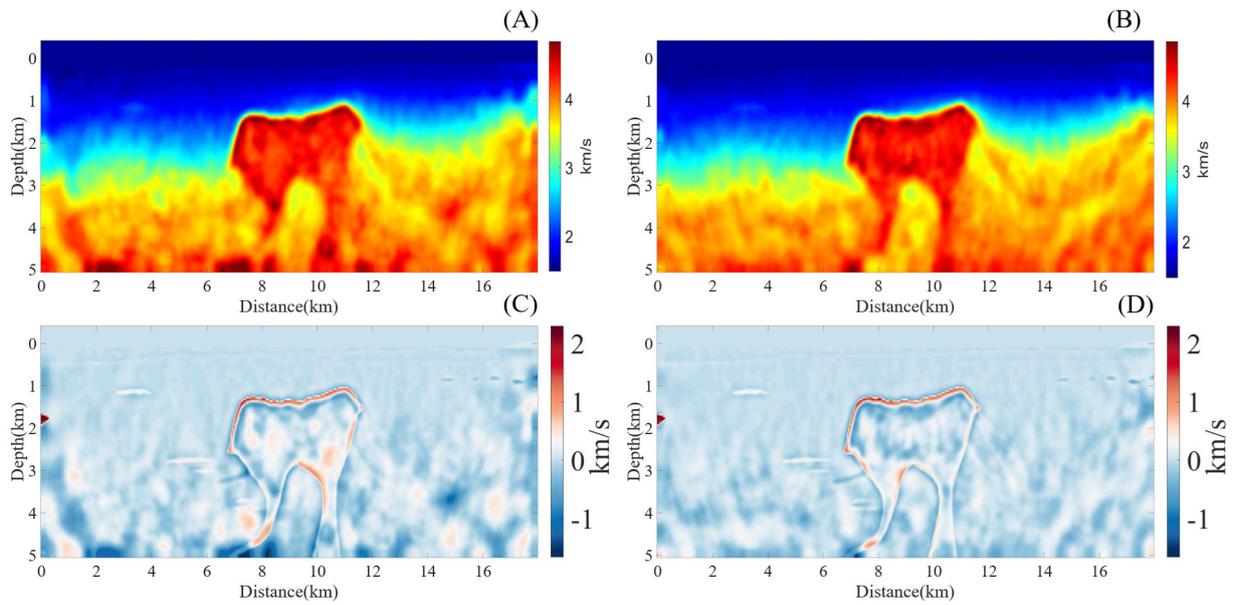

**Figure 10.** 2004 BP model. (A) Inversion result of Tikhonov regularization FWI. (B) Inversion result of the K-support norm FWI. Both were added 4.5 *dB* noise. (C) Difference between Figure 10-A and the true benchmark. (D) Difference between Figure 10-B and the true benchmark. The 4.5 *dB* random noise is shown in Figure 9.

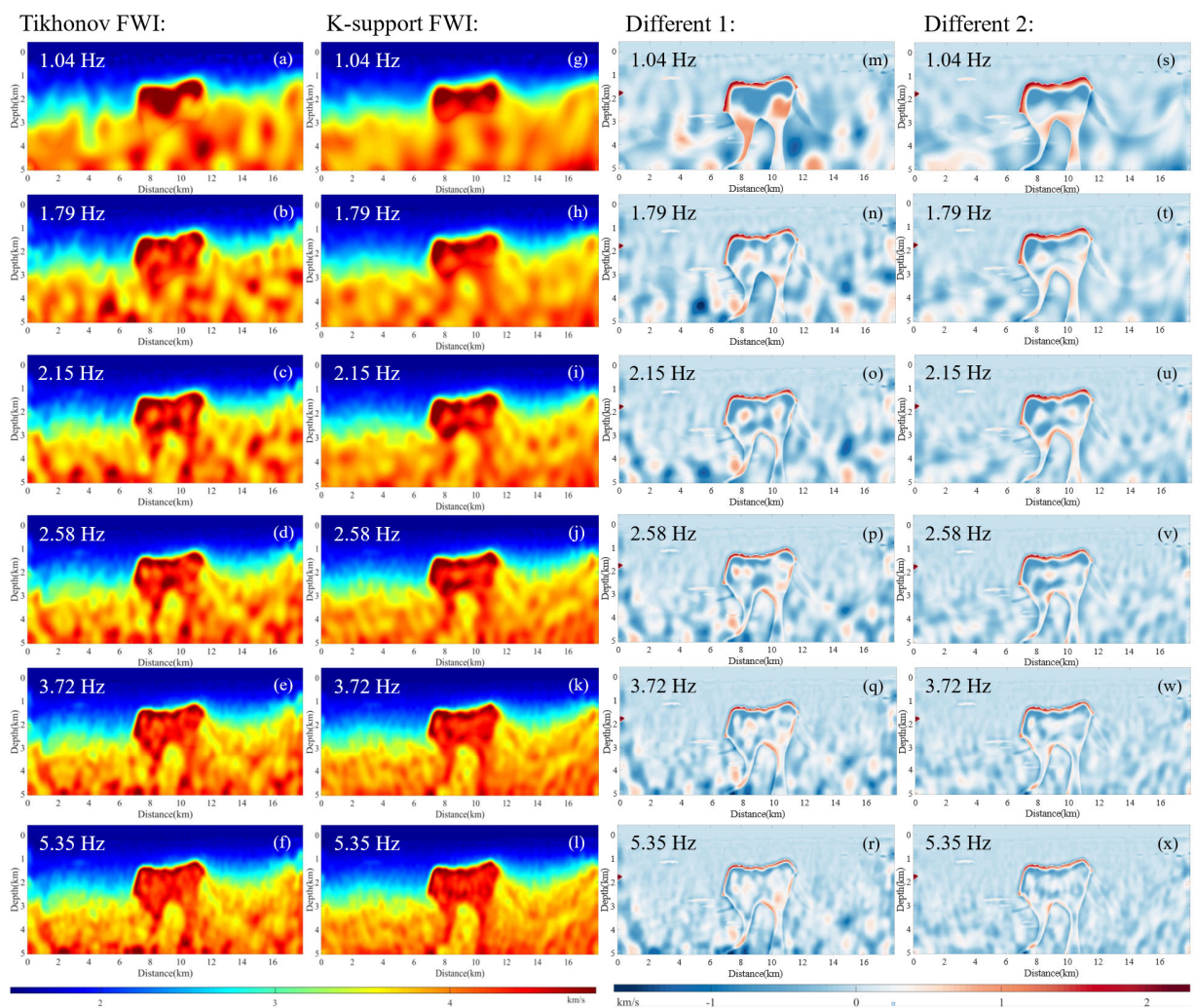

**Figure 11.** 2004 BP model. (a-f) Tikhonov regularization FWI inversion results in 1.04 Hz, 1.79 Hz, 2.15 Hz, 2.58 Hz, 3.72 Hz, and 5.35 Hz, respectively. (g-l) FWI based on K-support norm inversion results in 1.04 Hz, 1.79 Hz, 2.15 Hz, 2.58 Hz, 3.72 Hz, and 5.35 Hz, respectively.(m-r) Differences between the Tikhonov FWI and the true velocity model. (s-x) Differences between the K-support FWI and the true velocity model. It can be seen that sparsity plays an influential role in suppressing outliers and noise.

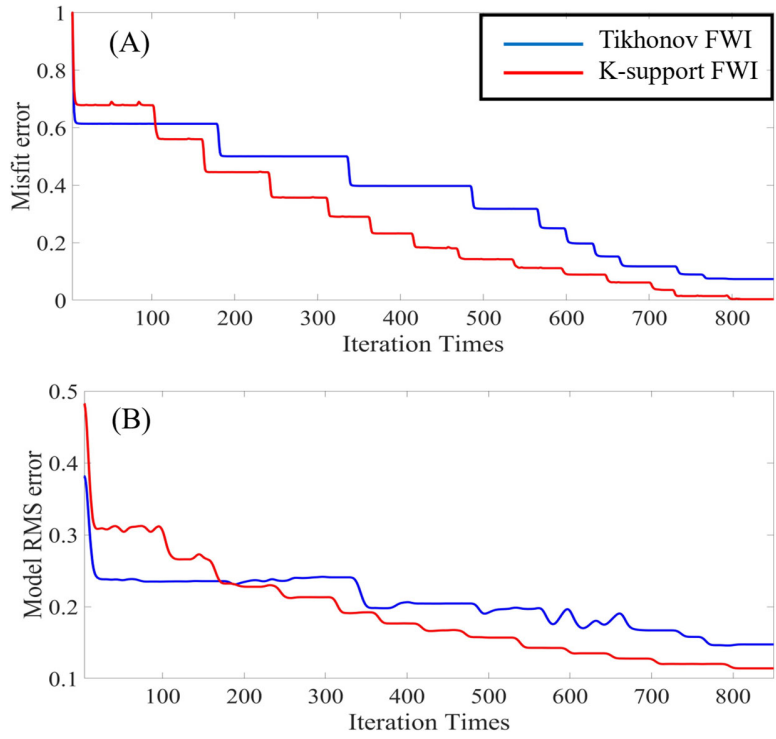

**Figure 12.** Model curves for the test. (A) Misfit error, (B) Model RMS error. The solid red line represents the K-support norm, and the blue line represents the Tikhonov regularization FWI. The formula of the Model RMS error is shown in Equation 11.

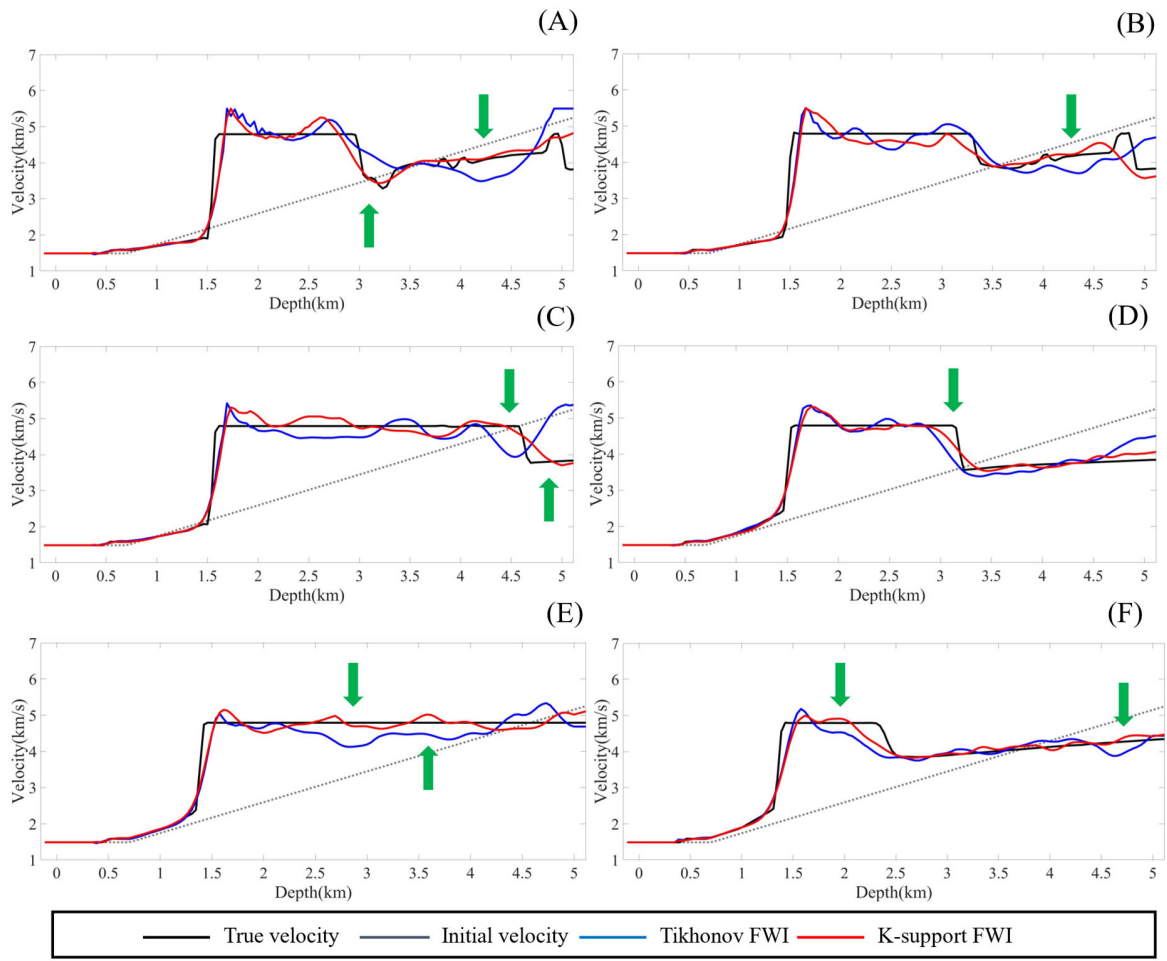

**Figure 13.** 2004 BP model.1-D velocity models at different *X*-positions. (A) *X*=7.4 km; (B) *X*=7.8 km; (C) *X*=8.2 km; (D) *X*=9.8 km;(E) *X*=10.4 km; (F) *X*=11.3 km. The vertical comparison of the actual velocity model (solid black line), initial velocity model (grey dotted line), the Tikhonov regularization FWI velocity model (solid blue line), and the K-support norm FWI velocity model (solid red line).

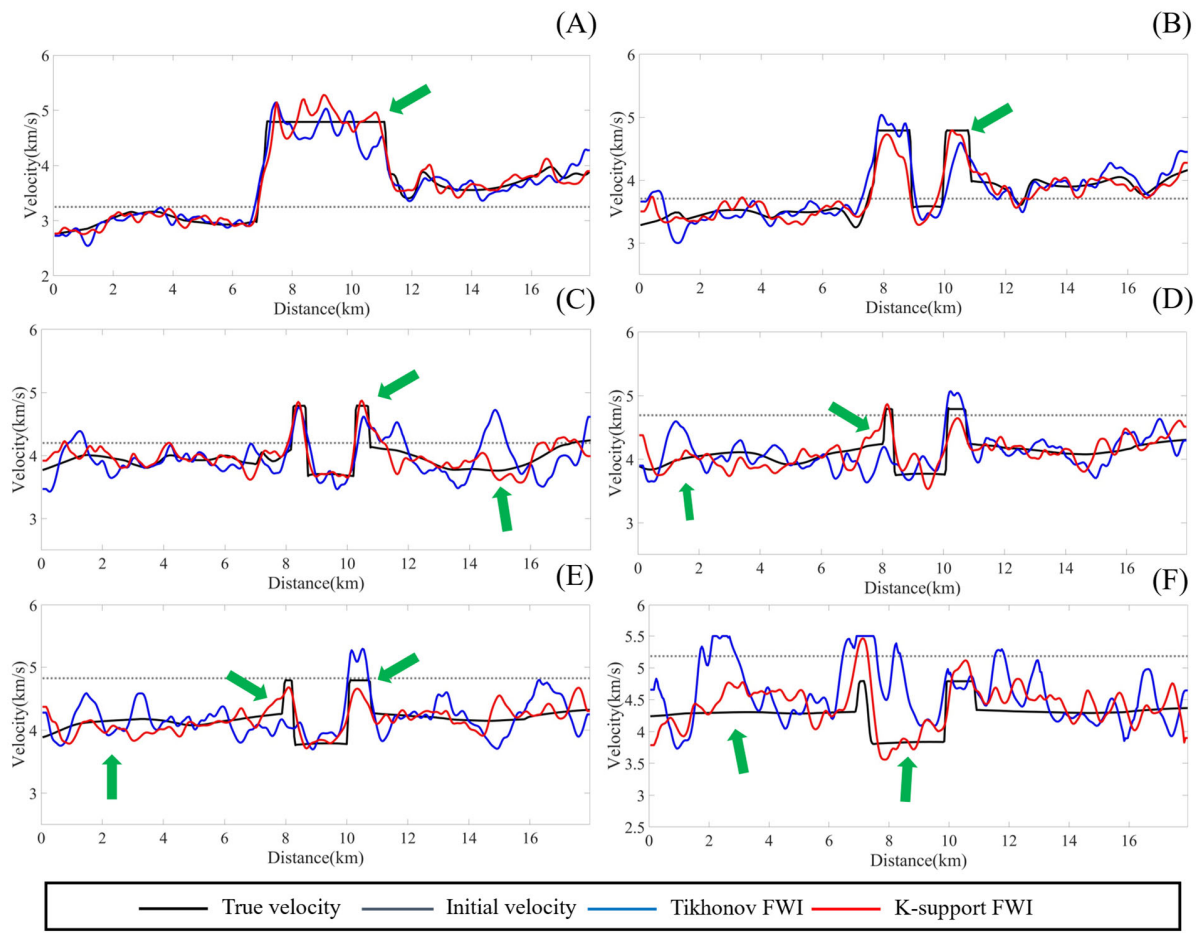

**Figure 14.** 2004 BP model. 1-D velocity models at different *Y*-positions. (A) *Y*=2.83 km; (B) *Y*=3.35 km; (C) *Y*=3.91 km; (D) *Y*=4.47 km; (E) *Y*=4.61 km; (F) *Y*=5.03 km. The lateral comparison of the actual velocity model (solid black line), initial velocity model (grey dotted line), the Tikhonov regularization FWI velocity model (solid blue line), and the K-support norm FWI velocity model (solid red line).

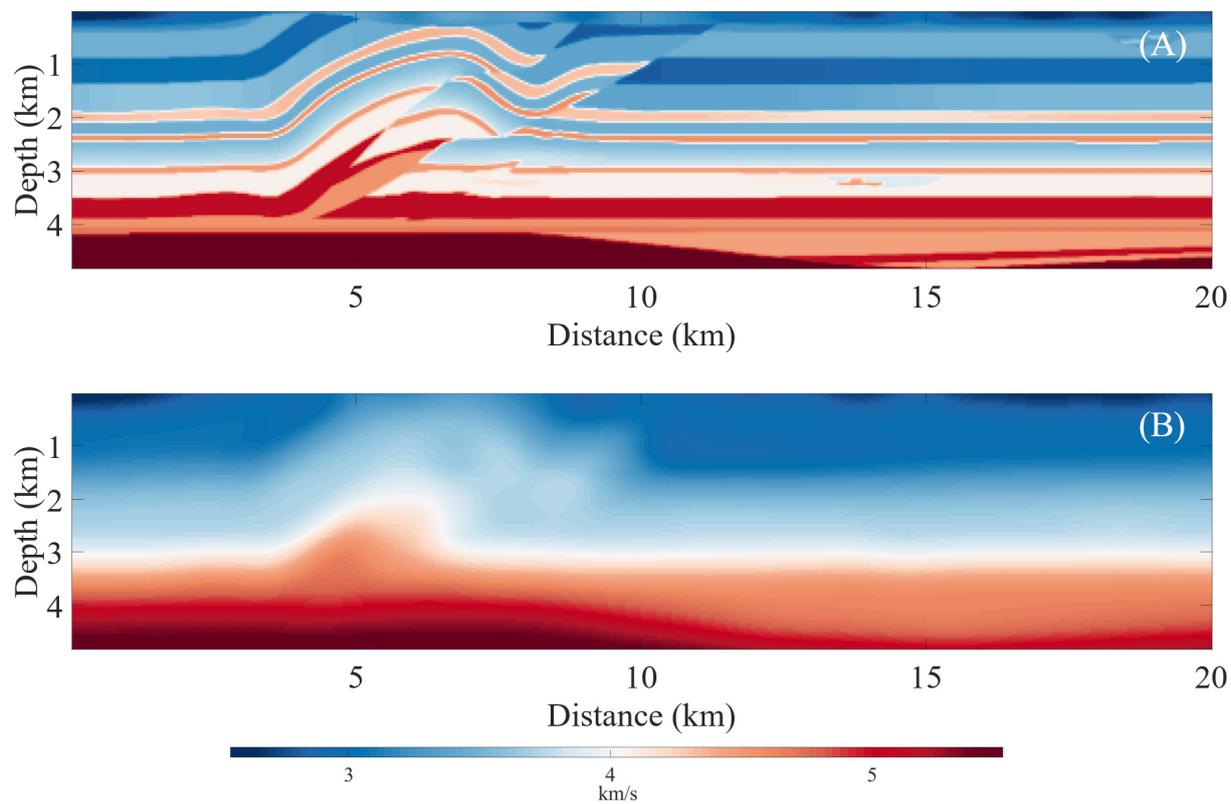

**Figure 15.** SEG/EAGE Overthrust Model. (A) True velocity model. (B) Initial velocity model.

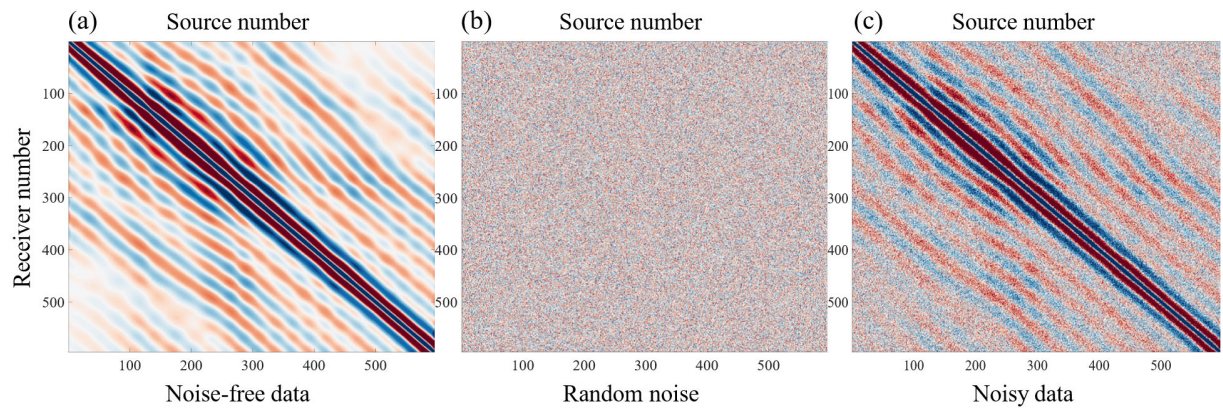

**Figure 16.** The real part of the 3.22 Hz data for the SEG/EAGE Overthrust model. (a) Noise-free data; (b) Random noise; (c) Noisy data.

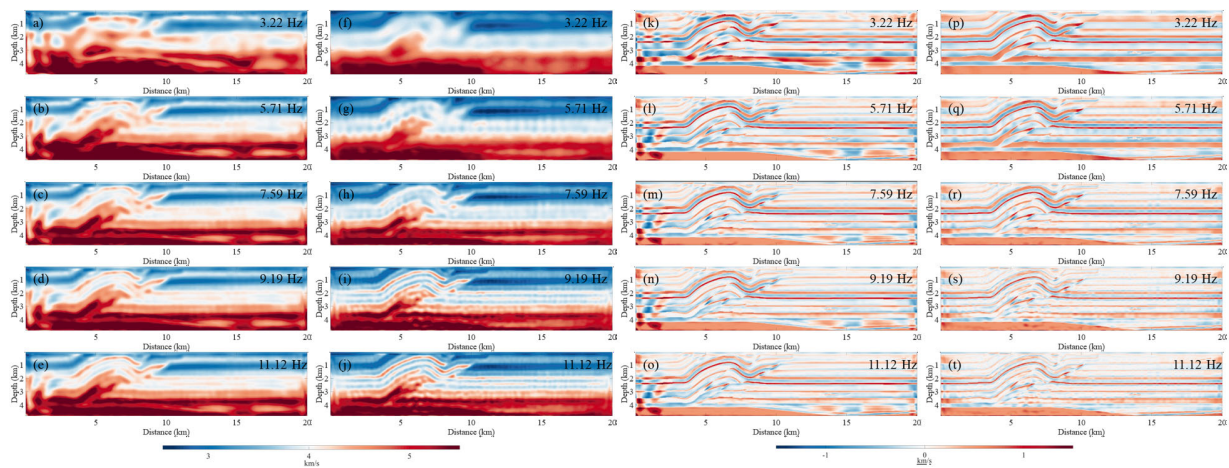

**Figure 17.** SEG/EAGE Overthrust Model. (a-e) Tikhonov regularization FWI inversion results in 3.22 Hz, 5.71 Hz, 7.59 Hz, 9.19 Hz, and 11.12 Hz, respectively. (f-j) FWI based on K-support norm inversion results in 3.22 Hz, 5.71 Hz, 7.59 Hz, 9.19 Hz, and 11.12 Hz, respectively. (k-o) Differences between the Tikhonov FWI and the true velocity model. (p-t) Differences between the K-support FWI and the true velocity model. Both were added to the random noise; the noise information is shown in Figure 16.

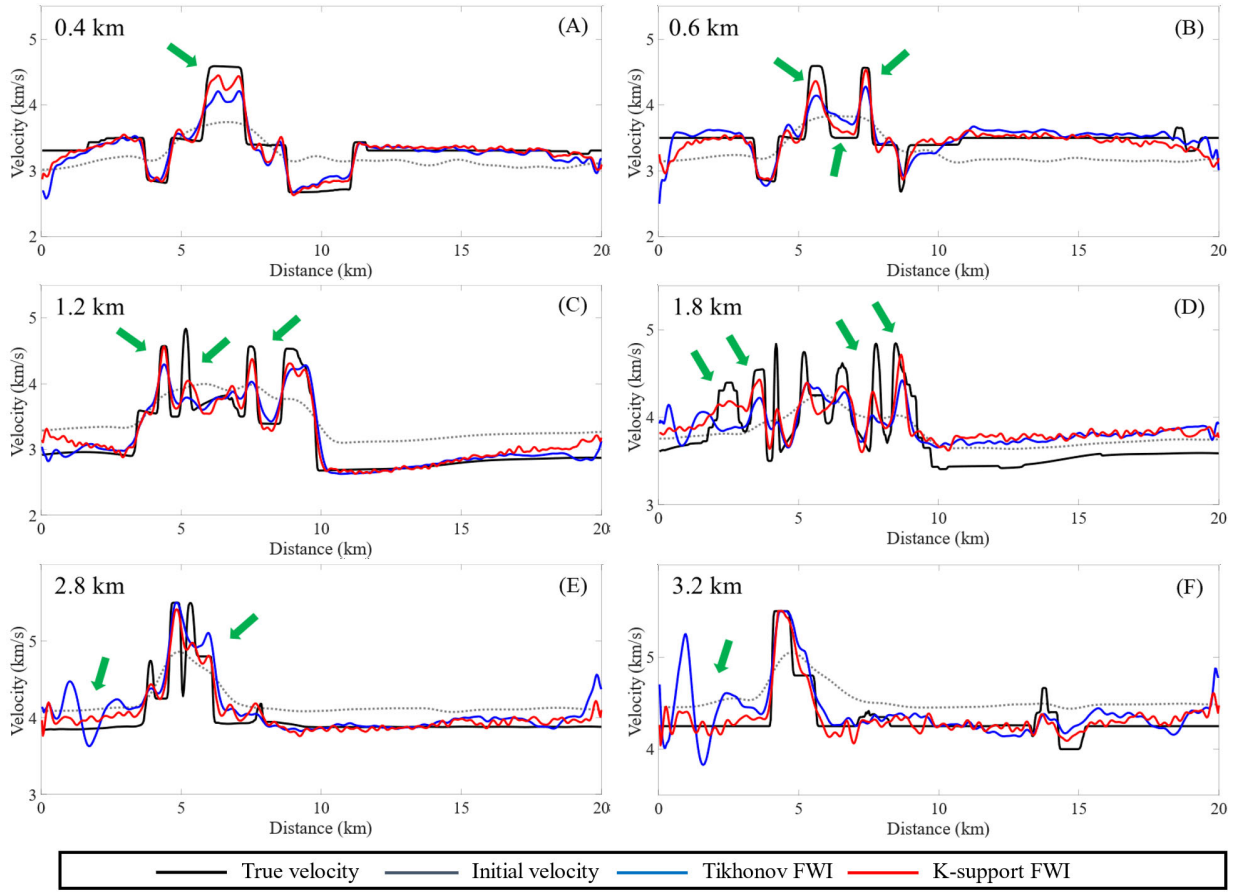

**Figure 18.** SEG/EAGE Overthrust Model, 1-D velocity models at different *Y*-positions. (A) *Y* = 0.4 km; (B) *Y* = 0.6 km; (C) *Y* = 1.2 km; (D) *Y* = 1.8 km; (E) *Y* = 2.8 km; (F) *Y* = 3.2 km. The lateral comparison of the actual velocity model (solid black line), initial velocity model (grey dotted line), the Tikhonov regularization FWI velocity model (solid blue line), and the K-support norm FWI velocity model (solid red line).

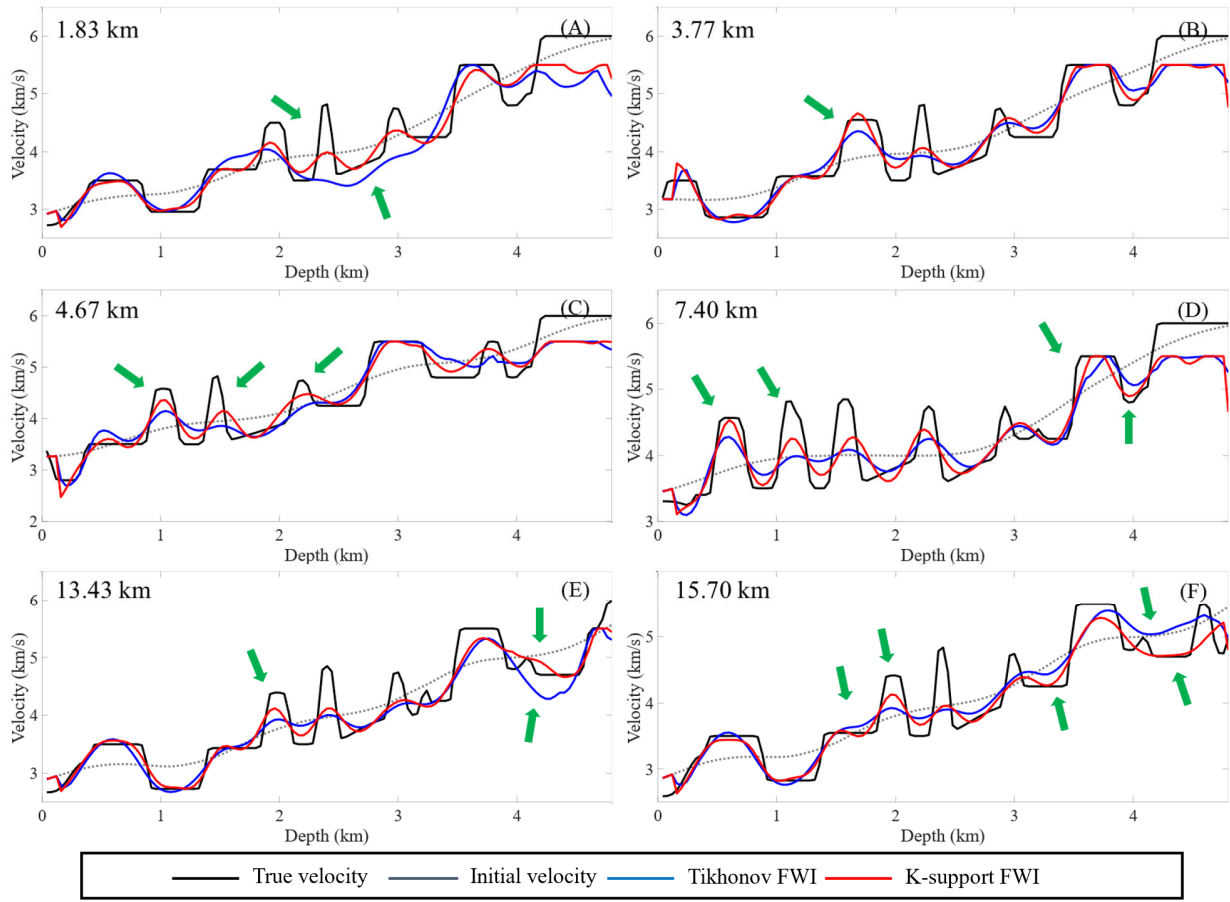

**Figure 19.** SEG/EAGE Overthrust Model, 1-D velocity models at different *X*-positions. (A) *X* = 1.83 km; (B) *X* = 3.77 km; (C) *X* = 4.67 km; (D) *X* = 7.40 km; (E) *X* = 13.43 km; (F) *X* = 15.70 km. The vertical comparison of the actual velocity model (solid black line), initial velocity model (grey dotted line), the Tikhonov regularization FWI velocity model (solid blue line), and the K-support norm FWI velocity model (solid red line).

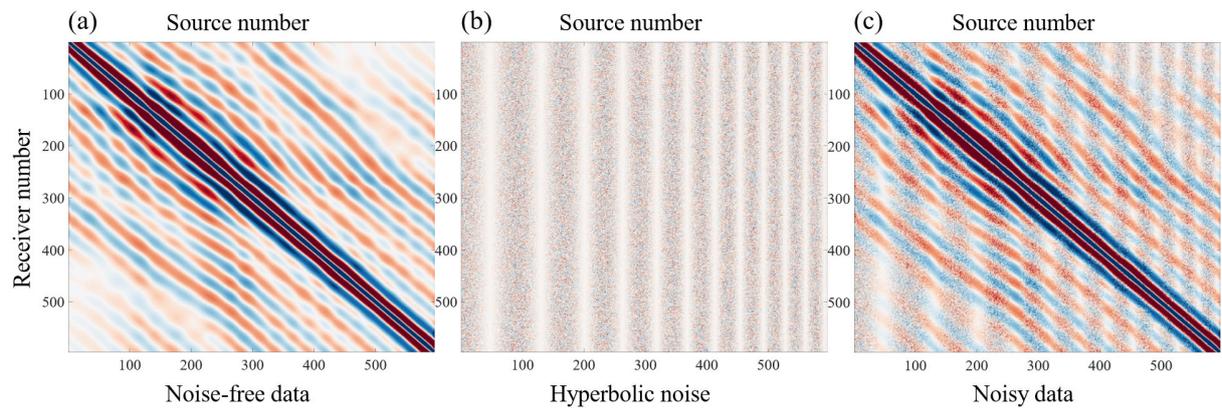

**Figure 20.** The real part of the 3.22 Hz data for the SEG/EAGE Overthrust model. (a) Noise-free data; (b) Hyperbolic noise; (c) Noisy data.

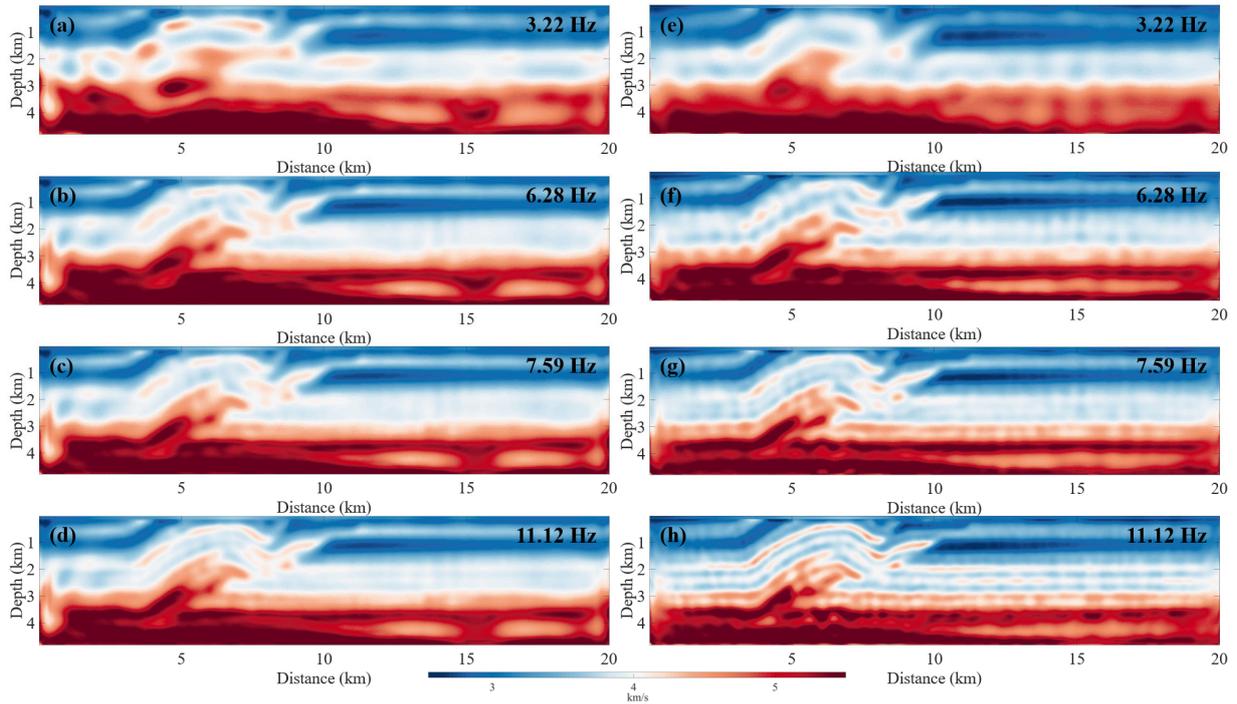

**Figure 21.** SEG/EAGE Overthrust Model. (a-d) Tikhonov regularization FWI inversion results in 3.22 Hz, 6.28 Hz, 7.59 Hz, and 11.12 Hz, respectively. (e-h) FWI based on K-support norm inversion results in 3.22 Hz, 6.28 Hz, 7.59 Hz, and 11.12 Hz, respectively. Both were added to the hyperbolic noise; the noise information is shown in Figure 20.